\documentclass[]{aastex}
\usepackage{emulateapj5}
\usepackage{onecolfloat}
\usepackage{graphicx} 
\usepackage{fancyheadings} 
\usepackage{ulem}
\usepackage{rotating}
\usepackage{lscape}

\newcommand{\delv}{\Delta v}

\newcommand{\lya}{Ly$\alpha$ }

\newcommand{\cm}[1]{\, {\rm cm^{#1}}}
\newcommand{\N}[1]{{N({\rm #1})}}

\newcommand{\sci}[1]{{\rm \; \times \; 10^{#1}}}

\newcommand{\cmma}{\;\;\; ,}

\newcommand{\mkms}{{\rm \; km\;s^{-1}}}

\begin{document}

\twocolumn[%
\submitted{Accepted to PASP: July 19, 2002}

\title{Probing the H\,I Kinematics of the LMC: 
Toward Interpreting QAL Observations of Protogalactic Velocity Fields}

\author{JASON X. PROCHASKA\altaffilmark{1,2}}
\affil{The Observatories of the Carnegie Institute of Washington}
\affil{813 Santa Barbara St. \\
Pasadena, CA 91101}
\email{xavier@ociw.edu}
\and
\author{EMMA RYAN-WEBER}
\affil{School of Physics; University of Melbourne\\
VIC 3010, Australia}
\email{eryan@physics.unimelb.edu.au}
\and
\author{LISTER STAVELEY-SMITH}
\affil{Australia Telescope National Facility; CSIRO\\
P.O. Box 76; Epping, NSW 1710, Australia}
\email{Lister.Staveley-Smith@csiro.au}

\begin{abstract} 

We examine the gas kinematics of the LMC revealed by the high
spatial and velocity resolution H\,I data cube of the combined
ATCA and Parkes telescope surveys.
We adopt an approach designed to facilitate comparisons with 
quasar absorption line observations, in particular restricting
our analysis to pointings with H\,I column density satisfying the
damped \lya (DLA) criterion.  We measure velocity widths for $\approx 5000$
random pointings to the LMC and find a median value of $\approx 40 \mkms$
which modestly exceeds the value predicted by differential rotation. 
This median is significantly lower than the median value observed for the
metal-line profiles of high $z$ DLA. 
Therefore,
assuming the metal-line profiles track H\,I kinematics,
the velocity fields of high $z$ DLA are inconsistent with the kinematics
of low-mass galaxies like the LMC.
We also investigate the kinematic characteristics of the giant
H\,I shells
which permeate the LMC.  These shells impart an additional 10 to 20~km/s
to the velocity widths in $\approx 20\%$ of random pointings to the LMC.
These non-gravitational motions are insufficient to explain
the DLA kinematics even if the shells had significanlty larger
covering fraction in the early universe.
To account for the DLA with processes related to SN
feedback requires winds with a qualitatively different nature than those
observed in the LMC.

\end{abstract}

]

\pagestyle{fancyplain}
\lhead[\fancyplain{}{\thepage}]{\fancyplain{}{PROCHASKA, RYAN-WEBER, \& STAVELEY-SMITH}}
\rhead[\fancyplain{}{PROBING THE H\,I KINEMATICS OF THE LMC:
}]{\fancyplain{}{\thepage}}
\setlength{\headrulewidth=0pt}
\cfoot{}

\altaffiltext{1}{Hubble Fellow}
\altaffiltext{2}{Current Address: UCO/Lick Observatory; UC Santa Cruz;
Santa Cruz, CA 95064; xavier@ucolick.org}

\section{INTRODUCTION}
\label{sec-intro}

The introduction of echelle spectrographs on 10m class telescopes
has revealed the velocity fields of gas at high $z$
with unprecedented precision.  In the low density \lya forest, for
example, observations have investigated the thermal history and physical 
nature of the gas comprising the intergalactic medium 
\citep[e.g.][]{kirkman97,rauch98,bryan00,theuns02}.  
Similarly, studies of optically thick
absorbers probe the velocity fields of gas in non-linear,
collapsed or collapsing structures
\citep[hereafter PW97]{charlton96,rauch97,pw97}.  
Together these observations
constrain the nature of galaxy formation and examine
the processes of structure formation in the early universe.

Although every quasar sightline is rich in detail, 
their 'pinhole' nature complicates interpretation.
Quasar sightlines produce a core-sample
of an absorption system with data acquired along a single dimension
and with resolution along the redshift axis only.  
In general, one must assume a specific geometry 
in order to impose a spatial delineation to the observations.
In a few rare cases, 
lensed quasars or quasar pairs allow a multi-dimensional
analysis \citep[e.g.][]{dinshaw95,lopez99,rauch99,dodorico02}, but even
these observations are difficult to unravel.
In turn, one frequently finds that several models
can be introduced to explain specific sets of observations.

A good example of this degeneracy is found in
the damped \lya systems (DLA), quasar absorption line (QAL) systems
believed to represent the progenitors of present galaxies 
\citep{kauff96,steinmetz01}.  
PW97 and \cite{pw98} analysed the velocity fields of over 30 DLA
and ruled out several plausible morphologies for these high $z$ galaxies,
but their true physical nature remains uncertain.  
The DLA low-ion\footnote{The term 'low-ion' refers to the dominant ionization
state of an element in an H\,I gas (e.g.\ Fe$^+$, Si$^+$).} 
kinematics are remarkably well described by
a thick, rotating disk (PW97) but also by CDM scenarios 
involving the merging of multiple 'clumps' bound to
individual dark matter halos \citep{hae98,mcd99,maller01}.
Furthermore,
several authors have claimed the damped \lya systems might be
explained by outflows from SN winds \citep{nulsen98,schaye01}. 
Owing to the 'pinhole' nature of QAL studies and the absence of spatial
resolution, it is difficult to distinguish between these very different 
models.

One approach toward resolving the ambiguities of QAL research is to
repeat these experiments on present-day galaxies whose physical properties
and gas kinematics are well understood.  
Comparisons with velocity profiles obtained from high $z$ protogalaxies
could provide fresh insight into the physical nature of young galaxies.
Regarding the damped \lya systems, any galaxy with large
H\,I surface density is an analog. 
Surveys of H\,I-selected galaxies suggest the damped \lya systems
should exhibit a range of luminosity including both dwarf galaxies
and large spiral galaxies \citep{zwaan99,zwaan02,ryan02,rosenberg02}.
This includes the Milky Way,
the Magellanic Clouds, and many other Local Group galaxies.  
The Milky Way aside, it is difficult to probe even a single absorption
sightline through these galaxies because present technology limits UV
spectroscopy to magnitudes $V<16$, restricting the number of potential
background
targets to $\approx 0.01$ per square degree \citep[e.g.][]{shull02}.  
Without a major advance 
in UV space telescopes, one must consider alternative approaches.

In this paper we study the H\,I velocity fields of the Large
Magellanic Cloud (LMC).  
Optically, the LMC has a bar but otherwise has a chaotic appearance 
being peppered with bright H\,II regions. A stellar color-magnitude
study by \cite{smecker02} indicates that the LMC bar was formed
in a major episode around 5~Gyr ago, whilst the disk component of the LMC
was more gradually formed. In some respects, the LMC may have similar
properties
to the compact galaxies seen in the Hubble deep field \citep{colley97}.
The smooth disk component of the LMC is more easily seen in  the 2MASS/DENIS 
infrared study of \cite{vdmarel01}. 
In terms of gas kinematics, 
the ATCA HI study of \cite{kim98b} 
reveals the disk to be in regular rotation, though many sightlines
in the eastern half display multiple velocity components. Deviations from
regular rotation appear to be a result of star formation activity \citep{kim99},
tidal forces \citep{weinberg00} and, possibly, interaction with the halo
of the Milky Way \citep{deboer98}.

The following analysis 
examines the high resolution 21cm data cubes constructed from the
combined ATCA survey of \cite{kim98a} and the Parkes telescope
survey of \cite{staveley02}.
We perform a statistical investigation designed to facilitate comparisons
with QAL observations and  
shed new light on their interpretation,
in particular those of the damped \lya systems.
Furthermore, our analysis examines the kinematic impact of the 
H\,I shells which permeate the LMC \citep[e.g.][]{kim99}.  
We assess the contribution of their non-gravitational motions to the
observed velocity fields and consider the implications for interpretations
of the damped \lya kinematics.  Finally, we propose 
future observations which will build upon our treatment.

\section{ANALYSIS OF THE LMC}
\label{sec-anly}

The LMC HI data cube is from the combined 
Australia Telescope Compact Array (ATCA) survey of \cite{kim98a} and
the Parkes telescope survey of \cite{staveley02}. The ATCA
survey alone has a resolution of 60\arcsec, corresponding to 15~pc at the
distance of the LMC. However, being an interferometer, the ATCA is
insensitive
to structure with a spatial scale greater than sampled by its shortest
baseline, and the column density to each sightline will be
systematically low. Therefore the merged ATCA and Parkes data cube has been
used. Data from the merged cube has been presented in \cite{kim01} and
\cite{elmegreen01} and will be more fully presented
in Kim et al. (2002, in preparation).  The combined data cube
covers an area
of 10\arcdeg\ by 12\arcdeg, which includes and extends well beyond the main 
disk of the LMC. The velocity coverage is from 190 to 387 km s$^{-1}$ with
a resolution of 1.6 km s$^{-1}$. The brightness temperature sensitivity 
is 2.5 K which, for a linewidth of 40 km s$^{-1}$, corresponds to a 
column density sensitivity of $4 \sci{19} \cm{-2}$.

\begin{figure*}
\begin{center}
\includegraphics[height=6.0in, width=4.1in, angle=-90]{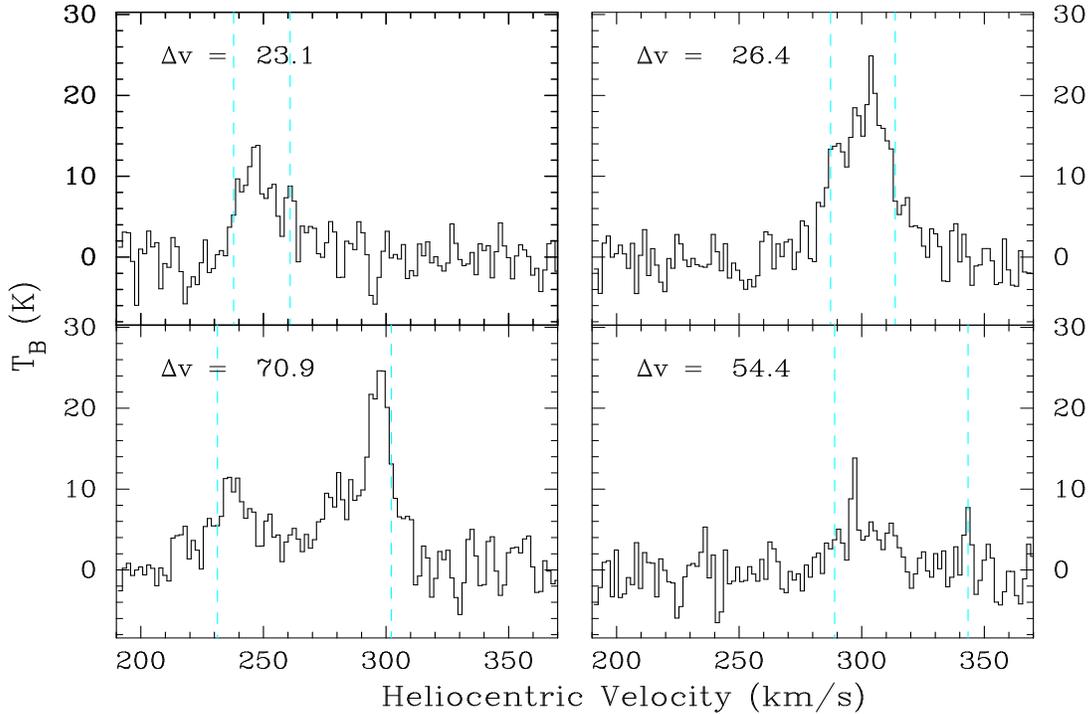}
\caption{Four representative pointings toward the LMC derived from the
combined ATCA and Parkes H\,I data sets.
The pointings are restricted to have
H\,I column density $\N{HI} \geq 2\sci{20} \cm{-2}$.  
Each pixel corresponds to a velocity
channel with width $\delta v = 1.6 \mkms$.  The brightness temperature
sensitivity is 2.5~K corresponding to a H\,I sensitivity of 
$4 \sci{19} \cm{-2}$ for a linewidth of 40~km/s.
The vertical dashed-lines indicate the velocity width $\delv$ which 
encompasses 90$\%$ of the integrated brightness temperature.
}
\label{fig:ex}
\end{center}
\end{figure*}

Our principal goal in this paper is to characterize the velocity fields
of the LMC in a fashion analogous to studies of quasar absorption line
systems, in particular the damped \lya systems.  The latter observations
focus on optically thin metal-line profiles
derived from 'pinhole' sightlines ($\lesssim 0.3$~pc wide) through
gas clouds with H\,I column density, $\N{HI} \geq 2 \sci{20} \cm{-2}$.
Because the absorbing galaxy is unrelated to the background quasar,
the sightline penetrates the galaxy at a random impact parameter 
selected according
to the cross-sectional distribution of H\,I gas.  
Ideally, one would probe the LMC in
an identical fashion but currently we cannot perform
this experiment.
Instead, we have examined
random 21\,cm pointings to the LMC through the 10\arcdeg\ by 12\arcdeg\ region
defining our combined H\,I data cube.  
For each pointing, we calculated the H\,I column density, 
\begin{equation}
\N{HI} = 1.823 \sci{18} \int T_b \, dv \; \cm{-2} \cmma
\end{equation}
assuming the H\,I gas is optically thin \citep{spitzer78}.
The analysis was restricted to sightlines where
$\log \N{HI} \geq 20.3$~dex. 
The H\,I data cube has
higher velocity resolution than QAL studies but each pointing encompasses
a significantly wider beam ($\approx 15$~pc).  This difference in beam size
could be important.  
\cite{lauroesch98} and \cite{meyer99} have demonstrated that
some metal-line transitions (notably Ca\,II and Na\,I) exhibit
significant density variations on pc-scales.  
Similarly, \cite{rauch99} have noted several km s$^{-1}$ offsets between
the velocity profiles of sightlines with separations of order 10~pc.
Although these small scale variations will be lost within the beam of the LMC 
analysis, their magnitude is too small to significantly affect our
main conclusions.

Apart from the beam size, the only other
obvious difference between DLA studies and our treatment of the
LMC is the specific ion used to probe the gas kinematics.
In the LMC, we observe H\,I gas exclusively whereas the DLA analysis
focuses on low-ions like Si$^+$, Fe$^+$, and Ni$^+$.  
Regarding the LMC, it is possible that the distinction is unimportant.
If the LMC exhibits a nearly constant metallicity like the Galaxy
\citep[both along given sightlines and also among multiple integrated
sightlines:][]{spitzer93,meyer98} and one focuses on non-refractory
elements (e.g.\ S, Zn), then the LMC metal-line profiles would closely
track the H\,I gas.  Of course, it will be important to demonstrate
this empirically with future observations of quasars behind the LMC.
The situation is more difficult for the damped \lya systems.
While the low-ion transitions for a given DLA all
tend to share very similar kinematics and relative
abundances \citep[PW97;][]{pro02b}, they might not
exactly trace the H\,I gas comprising the DLA.
Because observations of the saturated Lyman series do not resolve
the DLA H\,I kinematics, the correspondence
between metal-line and H\,I kinematics is unknown.
One could imagine, for example,
pockets of very low metallicity gas which would
contribute to an H\,I analysis but not appear in the metal-line
observations.  In this case, the metal-line profiles present a lower limit
to the magnitude of the velocity fields.  
Conversely, one could overestimate the kinematics of the
H\,I gas if there were regions along the sightline with very high
metallicity (e.g.\ a supernova shell).
Furthermore, if portions of the
metal-line profiles arise in ionized gas -- an unlikely prospect for
the majority of damped systems \citep{vladilo01,pro02a} -- the
metal-line profiles might overestimate the velocity fields of the H\,I gas.
Unfortunately, 
an examination of H\,I kinematics in the damped \lya systems awaits
the construction of a tremendous radio telescope.
We proceed in the hope that any differences are secondary
to the broad kinematic characteristics examined in this paper.

To compare the LMC kinematics with high redshift damped \lya systems,
we randomly selected 10000 pointings toward the LMC and analysed
those with $\N{HI} \geq 2 \sci{20} \cm{-2}$ ($\approx 50\%$).
Figure~\ref{fig:ex} presents velocity profiles for 4 representative
pointings to the LMC where $v=0 \mkms$ corresponds to the heliocentric
rest-frame. The kinematic profiles are relatively
simple; the majority of H\,I emission is confined to one or two Gaussian
features. 
The data at $v > 340 \mkms$ correspond to
a region of the data cube expected to contain no signal and, therefore, 
this spectra characterizes the noise of the H\,I profiles.  
To minimize the effects of this noise,
we process the data by smoothing over 5 channels ($\delta v \approx 8$~km/s)
before further analysis.

\begin{figure}[ht]
\begin{center}
\includegraphics[height=4.0in, width=3.5in,angle=-90]{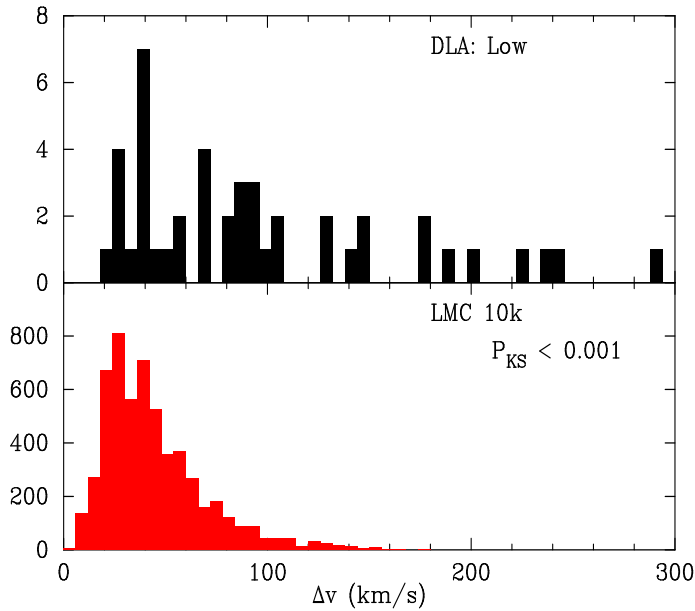}
\caption{
Velocity width distributions for $\approx 5000$ LMC random pointings
(bottom panel) compared against the $\delv$ distribution from the
high $z$ damped \lya systems (PW97; Prochaska \& Wolfe 1998; 
Prochaska \& Wolfe 2001).
A two-sided KS test reveals that the two distributions have a 
$<0.001$ probability
of having been drawn from the same parent population.  Therefore, the 
velocity fields of high $z$ protogalaxies are inconsistent with the 
gas kinematics of modern low-mass galaxies.
}
\label{fig:delv}
\end{center}
\end{figure}

Perhaps the most important kinematic characteristic of a velocity field
is its magnitude.  The standard means for assessing this magnitude is
through a measured velocity width $\delv$.
In the damped \lya systems, PW97
defined $\delv$ by the velocity interval which contains $90 \%$ of
the total optical depth.  
We apply the same statistic for the LMC,  as
marked by the vertical dashed-lines in Figure~\ref{fig:ex}. 
We have measured velocity widths for $\approx 5000$ random sightlines
toward the LMC.
Figure~\ref{fig:delv} compares this $\delv$ distribution against
the damped \lya sample \citep{pw98,pw01}.
The two samples have very different medians 
($\hat x_{LMC} = 38 \mkms, \hat x_{DLA} \approx 80 \mkms$)
and a two-sided Kolmogorov-Smirnov test indicates a $< 0.001$ probability
that they were drawn from the same parent population\footnote{The K-S test
is most sensitive to the median statistic of two distributions and is
insensitive to Poissonian noise \citep{press92}.  In the
following, one should interpret the K-S probability as a qualitative gauge
of the difference or similarity of two distributions.}.
The rotation speed of the LMC is 65~km/s assuming
an inclination angle of $33^\circ$ \citep{kim98b}; this implies
a maximum velocity width for differential rotation of 33~km/s.
Interestingly,
the majority of 'DLA-selected' pointings toward the LMC
exhibit larger widths. This indicates
the presence
of non-rotational motions and in the case of the largest velocity widths
(i.e.\ $\delv > 100 \mkms$) presumably non-gravitational velocity fields.
As noted above, these large $\delv$ values result from mechanisms 
including tidal forces \citep{weinberg00} and expanding H\,I shells
\citep{kim99}.

The contrast between the DLA and LMC velocity widths emphasizes one
of the principal conclusions drawn by \\
Prochaska and Wolfe:
{\it the gas kinematics of low-mass galaxies like the LMC are inconsistent
with the majority of damped \lya sightlines}.  
In particular, the results
obviate the key assumptions (e.g.\ rotation curve,
cloud-cloud velocity dispersion) imposed by Prochaska and Wolfe
to compare the DLA observations against models of low-mass galaxies.
Even at edge-on orientations,
the maximum velocity widths of the LMC would fall short of the 
median $\delv$ for the DLA.  
To explain the DLA $\delv$ distribution,
therefore, one requires much larger circular velocities within rotating disk
scenarios or an altogether different origin for the velocity fields
(e.g.\ winds, multiple clumps).
Perhaps the expanding H\,I shells observed within the LMC can account for
the DLA kinematics.  We consider this hypothesis in the following section.

\vskip 0.3in

\section{KINEMATICS OF H\,I SHELLS}
\label{sec-discuss}

H$\alpha$ emission maps of the LMC first revealed the presence of H\,I shells
throughout the galaxy \citep[e.g.][]{meaburn80}.  These shells and the holes they 
encompass are a generic feature of many nearby galaxies as cataloged 
through H\,I observations (e.g.\ M33: Deul \& den Hartog 1990; 
SMC: Staveley-Smith et al.\ 1997).  Their origin is linked to the combined
radiative and kinematic pressure of stellar winds from massive stars
and SN feedback \citep{weaver77}.  \cite{kim99} has extensively
surveyed the H\,I shells in the LMC and presented a classification
scheme which we adopt: giant shells are H\,I shells confined to the
main H\,I layer of the LMC (GS; $\ell < 360$~pc) and
supergiant shells have sizes which significantly exceed the H\,I
layer (SGS; $\ell > 360$~pc).  The shells have sizes ranging
from radii of 40~pc to 1.2~kpc following a power law with slope
$\alpha = -1.5 \pm 0.4$ over the interval 100 to 1000~pc.
The expansion velocity of the shells is well correlated with 
radius: $v_{exp} \approx 15 \mkms$ for the smallest shells
and $v_{exp} \approx 20-35 \mkms$ for the largest.

Of primary interest to the current analysis is addressing the impact of
these expanding H\,I shells on the gas kinematics of the LMC.
In terms of the DLA,
if gas-rich protogalaxies correspond primarily to dwarf galaxies
at high $z$, then one must introduce non-gravitational motions to explain
the larger velocity widths. Several authors have hypothesized
that SN winds, for example, may explain the DLA kinematics \citep{nulsen98}
and the velocity profiles observed in some Lyman limit systems
\citep{rauch99,bond01}.
In the LMC, we can examine the effects of SN feedback on the gas
kinematic characteristics.

\begin{figure*}
\begin{center}
\includegraphics[height=6.0in, width=4.0in, angle=-90]{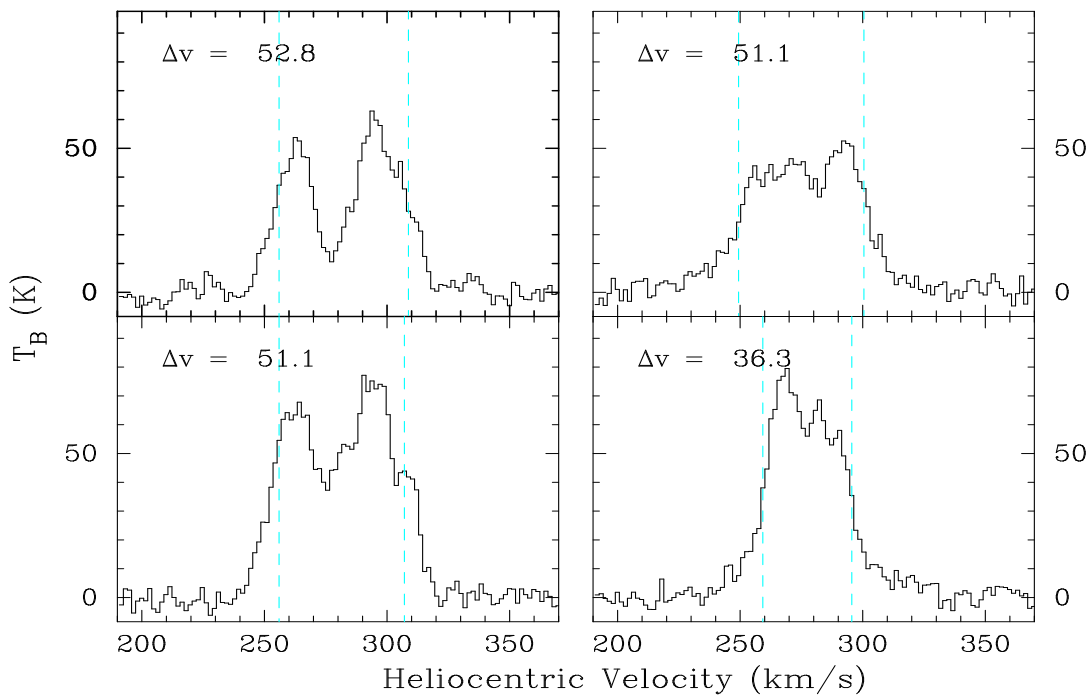}
\caption{
Representative pointings toward the expanding H\,I shell surrounding
30~Dor in the LMC.  Many profiles toward expanding shells exhibit the
double-peaked characteristic evident in the left-hand panels while the
remaining are similar to the profiles in the right-hand panels.
}
\label{fig:bubble}
\end{center}
\end{figure*}

\begin{figure*}
\begin{center}
\includegraphics[height=4.4in, width=4.0in]{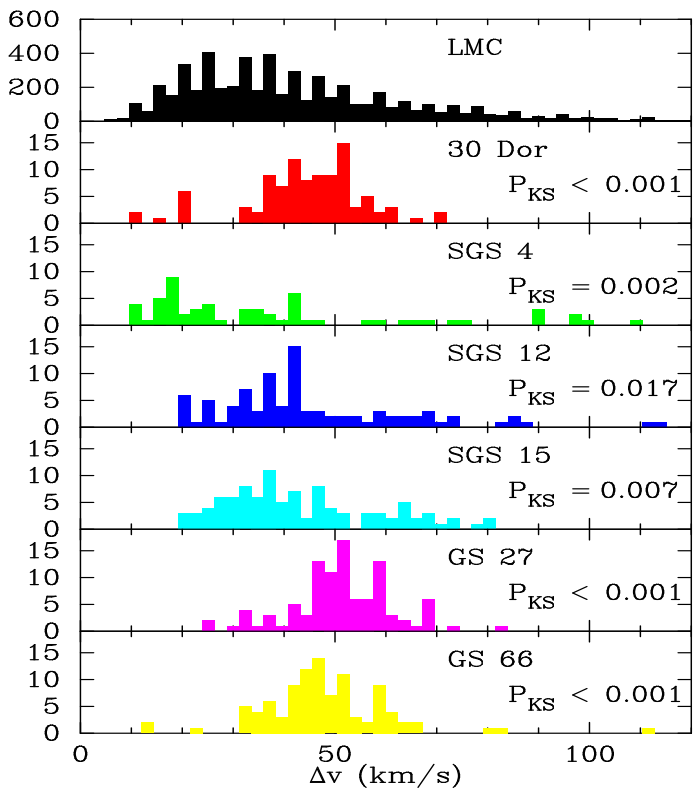}
\caption{
Velocity width distributions for the complete LMC sample (top panel) 
compared against random pointings to a series of H\,I shells identified 
by \cite{kim99}.  In general, the expanding H\,I shells exhibit 
systematically larger velocity widths, with median offsets of 
10 to 20~km/s.  The $P_{KS}$ probabilities indicate that these shells
exhibit significantly different kinematics than the typical LMC sightline.
}
\label{fig:lmcdelv}
\end{center}
\end{figure*}

We have investigated this aspect of the LMC velocity fields as follows.
We selected $\approx 300$ random pointings toward
several H\,I shells identified by \cite{kim99} and   
maintained the $\N{HI} = 2 \sci{20} \cm{-2}$ threshold.
We then measured the velocity width of each pointing with exactly
the same technique as the complete LMC sample.
Figure~\ref{fig:bubble} presents a representative sample of 4 sightlines
through the shell surrounding 30~Dor.  In the left panels, one observes
the double-peaked profiles characteristic of an expanding (or infalling)
shell.  A significant fraction of the sightlines share this characteristic.
The remaining profiles are more similar to the right
panels; these show nearly uniform optical depth or some mild asymmetry.
Figure~\ref{fig:lmcdelv} summarizes the velocity width distributions
for 6 H\,I shells.  For comparison, we plot the complete LMC $\delv$ distribution
(top panel).
The $P_{KS}$ values give the probability that the kinematic characteristics
of the gas comprising the H\,I shells are consistent with the general
LMC.  In most cases, the distributions are inconsistent with this
hypothesis at $< 0.01$ probability owing to the larger velocity widths
observed through the expanding H\,I shells.  Toward 30~Dor, for
example, the median 
$\delv$ value is $\approx 15 \mkms$ larger than the LMC sample.  

Our analysis demonstrates that the kinematics associated with expanding H\,I
shells can significantly influence the H\,I kinematics within low-mass
galaxies.  These non-gravitational velocity fields impart an additional
10 to 20~km/s to the velocity widths which corresponds to 
$33-50 \%$ of the median LMC velocity width.
At the same time, the significant difference between
the $\delv$ distributions of the H\,I shells and
the LMC sample indicate their filling factor and,
therefore, their overall kinematic impact is relatively small.
We found that the fraction of the H\,I data cube
covered by H\,I shells satisfying the DLA threshold is only $20 \%$.
This fraction is large enough to comprise most of the high $\delv$
tail observed in the LMC distribution, 
yet too small to grossly affect the LMC H\,I kinematics.

The results presented in Figure~\ref{fig:lmcdelv} suggest that the
H\,I shells produced via SN feedback are unlikely to explain the 
damped \lya kinematics.  Because the H\,I shells typically contribute
$< 20 \mkms$ to the observed velocity widths, these non-gravitational
motions would comprise only a small fraction of even the median
$\delv$ value
observed in the DLA.  Even if the porosity of these H\,I
shells was significantly larger in high $z$ dwarf galaxies, the H\,I kinematics
would still be inconsistent with the majority of damped systems.
As the porosity reached unity, however, the velocity and scale of the
shells might have a qualitatively different nature, i.e., a superwind.
Indeed, if SN feedback is to explain the DLA kinematics, the implied velocity
fields must have a very
different nature than the expanding H\,I shells observed in the LMC.
These superwinds may be more prevalent at high redshift owing to higher
star formation rates, feedback from 
Population III supernovae, or even differences in the relative
contribution of Type~Ia and Type~II SN.

\section{DISCUSSION}

Our simple analysis of the LMC's H\,I kinematics has revealed several
important results related to the dynamics of low-mass galaxies and, in turn,
the nature of high $z$ protogalaxies.  The median velocity width of a
sightline penetrating the LMC is $\approx 40 \mkms$ which modestly exceeds
the value predicted from differential rotation.  The offset must be
related to non-rotational motions and in many cases non-gravitational
velocity fields.  The latter are dominated by the expanding H\,I shells
which permeate the LMC \citep{kim99}.  Inspired by SN feedback, these shells
contribute an additional 10 to 20~km/s to the observed $\delv$ values.
For sightlines with large H\,I column densities,
the overall impact of the H\,I shells on the LMC kinematics is relatively
small.  We found that only $\approx 20\%$ of the random pointings 
intersect an H\,I shell.  This fraction is large enough, however, to play an
important role in the observed tail to large $\delv$ (Figure~\ref{fig:lmcdelv}).

The high resolution -- spatial and velocity --
of the combined H\,I data cube provides an excellent alternative to
QAL observations of the LMC.
By restricting our analysis to pointings 
$\N{HI} \geq 2 \sci{20} \cm{-2}$,  we compared the H\,I kinematic
characteristics of this low-mass galaxy with the velocity widths of
damped \lya metal-line profiles. 
The immediate and obvious
result (Figure~\ref{fig:delv}) is that the LMC exhibits a significantly
smaller-velocity field than the majority of damped systems, even when one
accounts for inclination.  
In short, low-mass galaxies with kinematics similar to the LMC
are inconsistent with high
$z$ DLA and, therefore, presumably the bulk of the protogalactic 
population in the early universe (e.g.\ Kauffmann 1996, Steinmetz 2001).
Similarly, the velocity fields of the LMC H\,I shells
are insufficient to explain the majority of DLA.  
The typical contribution to $\delv$ from these shells is only
10 to 20~km/s which is significantly lower than even the median DLA
velocity width.
Therefore, even if the
porosity of these bubbles approached unity at high $z$, the gas kinematics
expressed by the LMC H\,I shells could not explain the QAL observations.
Either the DLA galaxies have significantly higher mass than the LMC or
their velocity fields are dominated by very different physical
processes (e.g.\ superwinds, gravitational infall).

SN feedback has been introduced to explain the DLA kinematics
and other QAL phenomena \citep{nulsen98,rauch99,heckman00,bond01}.
As also emphasized by these authors,
our results indicate the winds and bubbles would be qualitatively
different from the H\,I shells observed in the LMC.
They must more closely resemble the outflows 
observed in starburst galaxies like NGC~520 \citep[e.g.][]{norman96} and
the $z\sim 3$ Lyman break galaxies \citep{steidel96}.
The winds must escape the galactic disk to (i) avoid stalling and (ii) achieve
large enough cross-section to account for a significant fraction of the DLA
\citep{schaye01}.
To explain the majority of large $\delv$ DLA, the
wind scenarios might require a duty cycle approaching 100$\%$ in dwarf galaxies
assuming the effects of random orientation and impact parameter lend
to a wide distribution of $\delv$ values.
If even $50 \%$ of the cross-section for DLA is made-up of quiescent 
galaxies, it would be very difficult to match the observed DLA kinematics.
There are several other pressing issues related to wind models.
\cite{nulsen98} predicted that DLA with large velocity width must also
exhibit high metallicity, yet several $z>3$ systems have [Fe/H]~$\approx -2$ and
$\delv > 150 \mkms$ \citep{pro01}.  
If their prediction applies to all outflow scenarios,
then another process must be introduced to explain this sub-sample of DLA.
Also, it is possible wind models would predict a greater fraction
of DLA with $\delv > 200 \mkms$ than observed, in particular if 
the same scenario is to
explain the Mg\,II absorption systems \citep[e.g.][]{bond01,rauch02}.
To address these concerns will require
the pursuit of high resolution
hydrodynamic simulations couched within the CDM framework.

In lieu of outflows, one might envision DLA
as the intersection of multiple, merging dwarf galaxies, each with kinematics
similar to the LMC \citep{hae98,maller01}.  There is, however, an 
important distinction: 
the H\,I radii of these merging dwarf galaxies must significantly exceed
that of the LMC and other modern dwarf galaxies in order
for QAL sightlines to routinely intersect more than one merging galaxy
\citep{maller01}.
Presently, these larger radii are not predicted by SPH simulations of 
galaxy formation \citep{gardner01,pw01}.  It remains to be seen whether
merging processes beneath the resolution of the SPH simulations
(e.g.\ tidal tails; Mihos \& Prochaska 2002, in preparation)
or SN feedback can satisfy this problem.

In conclusion, we wish to reemphasize the value of repeating the QAL
experiment on nearby, well-studied galaxies.  Recent quasar surveys toward
the LMC have borne several more bright targets which might permit observations
by HST \citep{geha02}.  Obviously, a comparison of the H\,I kinematics with
the absorption line profiles would be invaluable.  This line of
inquiry, however, will proceed very slowly until the next
generation of UV space telescopes.  Such an observatory would tremendously
impact our understanding of
high $z$ QAL systems and open the universe to similar inquiry at $z<1$.
Until then, one might pursue a 21~cm analysis 
for the SMC and possibly a few other Local Group galaxies.  
In particular, it would be extremely valuable to 
examine the gas kinematics of a starburst galaxy with a well established
superwind.

\acknowledgments

The authors wish to thank S. Kim for helping to provide the merged
H\,I data cube.  The authors acknowledge helpful conversations with
B. Weiner and E. Gawiser.
This work was partially supported by NASA through a Hubble Fellowship
grant HF-01142.01-A awarded by STScI to JXP.

\clearpage


\begin{thebibliography}{}

\bibitem[Bond et al.\ (2001)]{bond01}        
Bond, N.A., Churchill, C.W., Charlton, J.C., \& Vogt, S.S. 2001,
\apj, 562, 641

\bibitem[Bryan \& Machacek (2000)]{bryan00}  
Bryan, G.L. \& Machacek, M.E. 2000, \apj, 534, 57


\bibitem[Charlton \& Churchill (1996)]{charlton96}
Charlton, J.C. \& Churchill, C.W. 1996, \apj, 465, 631

\bibitem[Colley et al.\ (1997)]{colley97}     
Colley, W.N., Gnedin, O.Y., Ostriker, J.P., Rhoads, J.E. 
1997, \apj, 488, 579

\bibitem[de Boer et al.\ (1998)]{deboer98}  
de Boer, K. S., Braun, J. M., Vallenari, A., Mebold, U. 1998, \aap, 329, 49

\bibitem[Deul \& den Hartog (1990)]{deul90}  
Deul, E. \& den Hartog, R. 1990, \aap, 229, 362

\bibitem[Dinshaw et al.\ (1995)]{dinshaw95}   
Dinshaw, N., Foltz, C.B., Impey, C.D., Weymann, R.J., \&
Morris, S.L. 1995, \nat, 373, 223

\bibitem[D'Odorico, Petitjean, \& Cristiani (2002)]{dodorico02}   
D'Odorico, V., Petitjean, P., \& Cristiani, S. 2002, \aap, in press
(astro-ph/0205299)

\bibitem[Elmegreen, Kim, \& Staveley-Smith (2001)]{elmegreen01}
Elmegreen, B.G., Kim, S., \& Staveley-Smith, L. 2001, \apj, 548, 749

\bibitem[Gardner et al.\ (2001)]{gardner01}   
Gardner, J.P., Katz, N., Hernquist, L., \& Weinberg, D.H.
2001, \apj, 559, 131

\bibitem[Geha et al.\ (2002)]{geha02}  
Geha, M., et al.\, 2002, submitted

\bibitem[Haehnelt et al.\ (1998)]{hae98}
Haehnelt, M.G., Steinmetz, M. \& Rauch, M. 1998, \apj, 495, 647

\bibitem[Heckman et al.\ (2000)]{heckman00}     
Heckman, T.M., Lehnert, M.D., Strickland, D.K., \& Armus, L. 2000,
\apjs, 129, 493

\bibitem[Kauffmann (1996)]{kauff96}
Kauffmann, G. 1996, \mnras, 281, 475

\bibitem[Kim et al.\ (1998a)]{kim98a}         
Kim S., Staveley-Smith L., Dopita M. A., Freeman K. C., Sault R. J.,
  Kesteven M. J., McConnell D., 1998a, \apj, 503, 674

\bibitem[Kim et al.\ (1998b)]{kim98b}         
Kim, S., Staveley-Smith, L., Dopita, M.A., Freeman, K.C., 
Sault, R.J., Kesteven, M.J., \& McConnell, D. 1998b, \apj, 503, 729

\bibitem[Kim et al.\ (1999)]{kim99}         
Kim, S., Dopita, M.A., Staveley-Smith, L., \& Bessell, M.S. 1999, 
\aj, 118, 2797

\bibitem[Kim, Staveley-Smith, \& Sault (2001)]{kim01}         
Kim, S., Staveley-Smith, L., Sault, R. J. 2001, in Gas and Galaxy Evolution,
ASP Conference Proceedings, Vol. 240. Eds J. E. Hibbard, M. Rupen, 
J.H. van Gorkom (San Francisco: ASP), 435 

\bibitem[Kirkman \& Tytler (1997)]{kirkman97}   	
Kirkman, D. \& Tytler, D. 1997, \apj, 484, 672

\bibitem[Lauroesch et al.\ (1998)]{lauroesch98} 
Lauroesch, J.T., Meyer, D.M., Watson, J.K., \& Blades, J.C. 1998,
\apj, 507, 89L

\bibitem[Lopez et al.\ (1999)]{lopez99} 	
Lopez, S., Reimers, D., Rauch, M., Sargent, W.L.W., \& Smette, A.
1999, \apj, 513, 598


\bibitem[Maller et al.\ (2001)]{maller01}
Maller, A.H., Prochaska, J.X., Somerville, R.S., \& Primack, J.R.
2001, \mnras, 326, 1475

\bibitem[McDonald \& Miralda-Escud${\rm \acute e}$ (1999)]{mcd99}
McDonald, P. \& Miralda-Escud${\rm \acute e}$, J. 1999, \apj, 519, 486

\bibitem[Meaburn (1980)]{meaburn80}   
Meaburn, J. 1980, \mnras, 192, 365

\bibitem[Meyer, Jura, \& Cardelli (1998)]{meyer98}  
Meyer, D.M., Jura, M., \& Cardelli, J.A. 1998, \apj, 493, 222

\bibitem[Meyer \& Lauroesch (1999)]{meyer99}	
Meyer, D.M. \& Lauroesch, J.T. 1999, \apj, 520, 103L

\bibitem[Norman et al.\ (1996)]{norman96}   
Norman, C.A., Bowen, D.V., Heckman, T., Blades, C., \& Danly, L. 1996,
\apj, 472, 73

\bibitem[Nulsen, Barcons, \& Fabian (1998)]{nulsen98}  
Nulsen, P.E.J., Barcons, X., \& Fabian, A.C. 1998, \mnras, 301, 168

\bibitem[Press et al.\ (1992)]{press92}
Press, W. H., Teukolsky, S.A., Vetterling, W.T., \&
Flannery, B.P. 1992, Numerical Recipes in FORTRAN,
(New York: Cambridge University Press)

\bibitem[Prochaska (2002)]{pro02b}  	
Prochaska, J.X. 2002, in prep

\bibitem[Prochaska \& Wolfe (1997)]{pw97}  		
Prochaska, J. X. \& Wolfe, A. M. 1997, \apj, 486, 73 (PW97)

\bibitem[Prochaska \& Wolfe (1998)]{pw98}
Prochaska, J.X. \& Wolfe, A.M. 1998, \apj, 507, 113  

\bibitem[Prochaska \& Wolfe (2001)]{pw01}	
Prochaska, J.X. \& Wolfe, A.M. 2001, \apj, 560, L33

\bibitem[Prochaska et al.\ (2001)]{pro01}  
Prochaska, J.X., Wolfe, A.M., Tytler, D., Burles, S.M., Cooke, J.,
Gawiser, E., Kirkman, D., O'Meara, J.M., \& Storrie-Lombardi, L.
2001, \apjs, 137, 21

\bibitem[Prochaska et al.\ (2002)]{pro02a}  	
Prochaska, J.X., Howk, J.C., O'Meara, J.M., Tytler, D., 
Wolfe, A.M., Kirkman, D., Lubin, D., \& Suzuki, N. 2002, \apj, 571, 693

\bibitem[Rauch, Haehnelt, \& Steinmetz (1997)]{rauch97}  
Rauch, M., Haehnelt, M.G., \& Steinmetz, M. 1997, 481, 601

\bibitem[Rauch (1998)]{rauch98}     
Rauch, M. 1998, \araa, 36, 267

\bibitem[Rauch, Sargent, \& Barlow (1999)]{rauch99} 
Rauch, M., Sargent, W.L.W., \& Barlow, T. 1999, \apj, 515, 500

\bibitem[Rauch et al.\ (2002)]{rauch02} 
Rauch, M., Sargent, W.L.W., Barlow, T., \& Simcoe, R. 2002, \apj, in press
(astro-ph./0204461)

\bibitem[Rosenberg \& Schneider (2002)]{rosenberg02}
Rosenberg, J.L. \& Schneider, S.E. 2002, \apj, submitted (astro-ph/0202216)

\bibitem[Ryan-Weber et al.\ (2002)]{ryan02}  
Ryan-Weber, E., Webster, R., Staveley-Smith, L. 2002, in 
Extragalactic Gas at Low Redshift,
ASP Conf.\ Series v. 254, eds. J.S. Mulchaey and J.T. Stocke,
(San Fransisco: ASP), 209

\bibitem[Schaye (2001)]{schaye01}       
Schaye, J. 2001, \apj, 559, 1L

\bibitem[Shull (2002)]{shull02}   
Shull, J.M. 2002, in Extragalactic Gas at Low Redshift,
ASP Conf.\ Series v. 254, eds. J.S. Mulchaey and J.T. Stocke,
(San Fransisco: ASP), 51

\bibitem[Smecker-Hane et al.\ (2002)]{smecker02}   
Smecker-Hane, T.A., Cole, A.A., Gallagher, J.S., Stetson, P.B. 2002, \apj,
566, 239

\bibitem[Spitzer (1978)]{spitzer78}		
Spitzer, L., Jr. 1978, {\it Physical Processes in the Interstellar
Medium} (Wiley: New York), p. 156

\bibitem[Spitzer and Fitzpatrick (1993)]{spitzer93} 
Spitzer, L., Jr., \& Fitzpatrick, E.L. 1993, \apj, 409, 299

\bibitem[Staveley-Smith et al.\ (1997)]{staveley97} 	
Staveley-Smith, L., Sault, R.J., Hatzidimitriou, D., 
Kesteven, M.J., \& McConnell, D. 1997, \mnras, 289, 225

\bibitem[Staveley-Smith et al.\ (2002)]{staveley02} 	
Staveley-Smith, L., Kim, S., Calabretta, M. R., Haynes, R. F.,
Kesteven, M. J. 2002, \mnras 

\bibitem[Steidel et al.\ (1996)]{steidel96}
Steidel, C.C., Giavalisco, M., Dickinson, M., \&
Adelberger, K.L. 1996, \apj, 112, 352

\bibitem[Steinmetz (2001)]{steinmetz01}    
Steinmetz, M. 2001, in Galaxy Disks and Disk Galaxies,
ASP Conf.\ Series v. 230,
eds.\ J.G. Funes, S.J. and E. Maria Corsini, 
(San Fransisco: ASP), 633

\bibitem[Theuns et al.\ (2002)]{theuns02}  
Theuns, T., Zaroubi, S., Kim, T.-S., Tzanavaris, P., \& 
Carswell, R.F. 2002, \mnras, 332, 367

\bibitem[van der Marel (2001)]{vdmarel01}   
van der Marel, R. P., 2001, \aj, 122, 1827

\bibitem[Vladilo et al.\ (2001)]{vladilo01}    
Vladilo, G., Centuri${\rm \acute o}$n, M., Bonifacio, P., \&
Howk, J.C. 2001, \apj, 557, 1007

\bibitem[Weaver et al.\ (1977)]{weaver77}   
Weaver, R., McCray,, R., Castor, J., Shapiro, P., \& Moore, R.
1977, \apj, 218, 377

\bibitem[Weinberg (2000)]{weinberg00}  
Weinberg M. D., 2000, \apj, 532, 922

\bibitem[Zwaan, Verheijen, \& Briggs (1999)]{zwaan99}  
Zwaan, M.A., Verheijen, M.A.W., \& Briggs, F.H. 1999, PASA, 16, 100

\bibitem[Zwaan, Briggs, \& Verheijen (2002)]{zwaan02}  
Zwaan, M.A., Briggs, F.H., \& Verheijen, M.A.W. 2002, 
in Extragalactic Gas at Low Redshift,
ASP Conf.\ Series v. 254, eds. J.S. Mulchaey and J.T. Stocke,
(San Fransisco: ASP), 169


\end{thebibliography}
\end{document}